\documentclass[prb,aps,twocolumn,showpacs,superscriptaddress]{revtex4}

\usepackage{amsmath,amssymb,amsthm}
\usepackage{graphicx}
\usepackage{subfigure}
\usepackage{amsmath}
\usepackage{amsfonts,amssymb}
\usepackage{bm}
\usepackage{float}
\usepackage{color}
\usepackage{sidecap}
\allowdisplaybreaks
\usepackage{soul}
\setstcolor{red}
\usepackage[normalem]{ulem}
\usepackage{hyperref}
\hypersetup{colorlinks=true,breaklinks,urlcolor=blue,linkcolor=blue,citecolor=blue}

\begin{document}

\title{Dual-mapping and quantum criticality in off-diagonal Aubry-Andr\'{e} models}
\author{Tong Liu}
\thanks{t6tong@njupt.edu.cn}
\affiliation{School of Science, Nanjing University of Posts and Telecommunications, Nanjing 210003, China}
\author{Xu Xia}
\thanks{1069485295@qq.com}
\affiliation{Chern Institute of Mathematics and LPMC, Nankai University, Tianjin 300071, China}

\date{\today}

\begin{abstract}
We study a class of off-diagonal quasiperiodic hopping models described by one-dimensional Su-Schrieffer-Heeger chain with quasiperiodic modulations. We unveil a general dual-mapping relation in parameter space of the dimerization strength $\lambda$ and the quasiperiodic modulation strength $V$, regardless of the specific details of the quasiperiodic modulation. Moreover, we demonstrated semi-analytically and numerically that under the specific quasiperiodic modulation, quantum criticality can emerge and persist in a wide parameter space.
These unusual properties provides a distinctive paradigm compared with the diagonal quasiperiodic systems.
\end{abstract}

\pacs{71.23.An, 71.23.Ft, 05.70.Jk}
\maketitle

\section{Introduction}
\label{n1}
Since the publication of Anderson's seminal paper~\cite{1an},
the metal-insulator transition has been studied in a wide range of quantum disordered systems.
According to the scaling theory~\cite{2scal}, there is no metal-insulator transition in one-dimensional (1D) systems with
random on-site potentials. However, the eigenstate near the band center in the off-diagonal random hopping
disorder~\cite{Fleishman,3Inui} exhibits the delocalization. This model has a chiral symmetry which is not present in the standard Anderson case with diagonal/on-site disorder. At the band center $E=0$, the conductance exhibits strong fluctuations and displays an
algebraically decaying mean value, and the density of state has a logarithmically divergent scaling behavior, which is different from the standard localized state~\cite{4Ziman}.

On the other hand, 1D quasiperiodic models~\cite{1PRB,Gramsch,mobility1,mobility2} which can host localized, extended or critical eigenstates also attracts much interest in view of its rich physics.
The Aubry-Andr\'{e} (AA) model~\cite{7aubry} is an important paradigm of 1D quasiperiodic systems.
It can be derived from the reduction of a two-dimensional quantum Hall system in the magnetic field. Due to recent advances in experimental techniques, the AA model has been realized in ultracold atoms~\cite{8BILLY,9ROATI} and photonic crystals~\cite{10PRL,11PRL}.
The phase diagram of the AA model has been well understood with extensive researches~\cite{LXJ1,LXJ2,LXJ3}, and many different variations of the AA model were studied~\cite{liu1,liu2}.

Beyond the diagonal AA model, the off-diagonal AA model may also exhibit an abundant physical phenomena.
The commensurate off-diagonal AA model can host the zero-energy topological edge modes~\cite{17PRL}, whereas the incommensurate cases~\cite{18PRB,19PRB} can bring the localization-delocalization phase transition. Inspired by the the off-diagonal random hopping model, we wondered that does it exist the novel physical phenomena in the off-diagonal quasiperiodic hopping model.

In this paper, we explore a class of off-diagonal quasiperiodic hopping models described by one-dimensional Su-Schrieffer-Heeger chain with quasiperiodic modulation, we unveil a general dual-mapping relation in the parameter space. Moreover we discover there exists a wide critical region under the specific quasiperiodic modulation. Critical states, also known as multifractal states, i.e., the non-ergodic extended states do exist at the phase transition point of 3D Anderson model and 1D AA model. The emergence of the concept of many-body localization
(MBL) inspires enormous interest of searching the non-ergodic phase. However, due to the complexity of many-body
systems it is worthwhile to explore and demonstrate existence of the non-ergodic phase of the single-particle system in a wide parameter space (not only a phase transition point). References~\cite{Luca,Alt} have numerically demonstrate such phase exists in a finite range of the disorder strength on the random regular graph (a treelike graph without boundary). Our work provides another paradigm which possesses this distinct phase in a wide parameter space.

\section{Model and Dual-mapping}
\label{n2}
\begin{figure}
	\centering
	\includegraphics[width=0.48
	\textwidth]{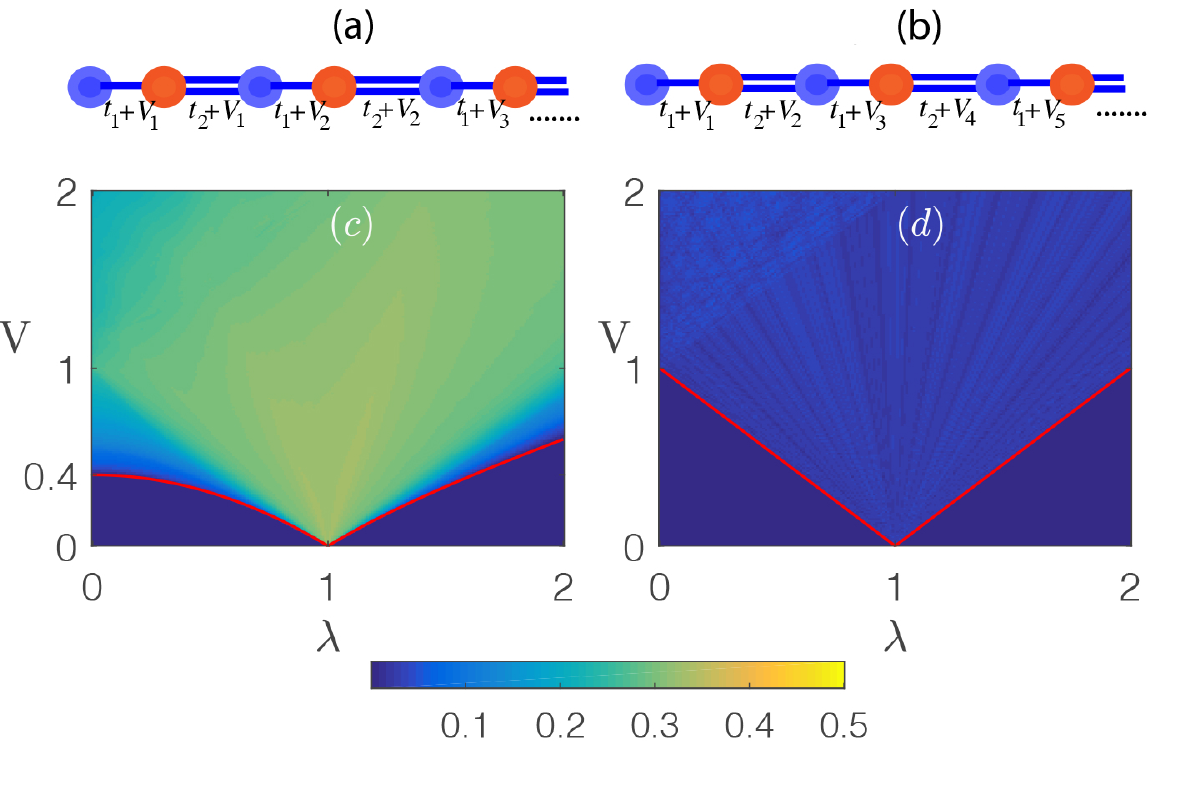}\\
	\caption{(Color online) (a) Schematic of a SSH chain comprising $N$ unit cells with the intra-hopping amplitude $t_1$, the inter-hopping amplitude $t_2$ and the quasiperiodic perturbation $\{\{V_{n-1},V_{n},V_{n},V_{n}\}$ (model I). (b) Schematic of a SSH chain comprising $N$ unit cells with the intra-hopping amplitude $t_1$, the inter-hopping amplitude $t_2$ and the quasiperiodic perturbation $\{V_{2n-2},V_{2n-1},V_{2n-1},V_{2n}\}$ (model II). (c) MIPR of model I in the ($\lambda,V$) parameter space. Two red lines represent two extended-nonextended phase transition lines $V=0.4(1-\lambda^2)$ ($0\leq\lambda\leq1$) and $V=0.4(\lambda-1/\lambda)$ ($1\leq\lambda$). (d) MIPR of model II in the ($\lambda,V$) parameter space. Two red lines represent two extended-nonextended phase transition lines $V=1-\lambda$ ($0\leq\lambda\leq1$) and $V=\lambda-1$ ($1\leq\lambda$). The total number of unit cells is set to be $L=1000$}
	\label{fig1}
\end{figure}
We consider a class of off-diagonal quasiperiodic hopping models, more intuitively, it can be viewed as a Su-Schrieffer-Heeger (SSH) chain with quasiperiodic modulation. The Hamiltonian of the model is expressed as
\begin{equation}\label{difference}
\begin{aligned}
&E a_n=(t_2+V_{m1})b_{n-1}+(t_1+V_{m2})b_{n},\\
&E b_n=(t_1+V_{m3})a_{n}+(t_2+V_{m4})a_{n+1},
\end{aligned}
\end{equation}
In a disorder-free lattice with intra- and inter-hopping amplitudes, the SSH lattice exhibits two
topologically distinct phases, characterized by a distinct winding number $Q=0$ for $t_1>t_2$ and $Q=1$ for $t_1<t_2$. Here we
consider a SSH chain in the topological nontrivial phase $t_1=1-\lambda$ and $t_2=1+\lambda$, where $\lambda$ is the dimerization strength.
$a_n$ ($b_n$) are the wave function on sublattice A (or B) of the $n$-th unit cell,
and there exists an extra quasiperiodic modulation $V_{n}=V\cos(2 \pi \alpha n+\theta)$. A typical choice of the parameters is $\alpha=(\sqrt{5}-1)/2$. For convenience, $t = 1$ is set as the energy unit.

It should be noted that, the quasiperiodic modulation in Eq.~(\ref{difference}) can take various sequences. Here, we choose two common sequences, $\{V_{m1},V_{m2},V_{m3},V_{m4}\}=\{V_{n-1},V_{n},V_{n},V_{n}\}$ (model I) illustrated in Fig.~\ref{fig1}(a) and $\{V_{m1},V_{m2},V_{m3},V_{m4}\}=\{V_{2n-2},V_{2n-1},V_{2n-1},V_{2n}\}$ (model II) illustrated in Fig.~\ref{fig1}(b).
The Hermitian nature of two systems still remain.

In order to investigate the localization nature of two models, we numerically solving Eq.~(\ref{difference}), and obtain two components $a_{n}$ and $b_{n}$ of the wave functions. The inverse participation ratio (IPR) is usually used to study the localization-delocalization transition~\cite{IPR21,IPR22}. For any given normalized wave function, the corresponding IPR is defined as
${\rm{IPR}}_n =\sum^{L}_{j=1}\left(\left|a_{n,j}\right|^{4}+\left|b_{n,j}\right|^{4}\right),$
which measures the inverse of the number of sites being occupied by particles. It is well known that the IPR of an extended state scales
like $L^{-1}$ which approaches zero in the thermodynamic limit. However, for a localized state, since only finite number of sites are
occupied, the IPR is finite even in the thermodynamic limit. The mean of IPR over all the $2L$ eigenstates is dubbed the MIPR which is
expressed as ${\rm{MIPR}}=\frac{1}{2L}\sum_{n=1}^{2L}{\rm{IPR}}_{n}.$

In Fig.~\ref{fig1} (c) and (d), we plot MIPR of model I and model II as a function of $\lambda$ and $V$, respectively.
The dark blue regions represents the low amplitudes of MIPR, correspond to the extended phase, other colour regions correspond to nonextended phase, and the extended-nonextended phase transition lines (red lines) can be determined numerically. Interestingly, the phase transition lines in regions $(0\leq\lambda\leq1)$ and $(1\leq\lambda)$ seems to be a connection, i.e., $V=0.4(1-\lambda^2)\leftrightarrow V=0.4(\lambda-1/\lambda)$ for model I, and $V=1-\lambda \leftrightarrow V=\lambda-1$ in model II. To explain this phenomena, here we unveil a dual-mapping relates two dark blue regions (extended phase). Eq.~(\ref{difference}) of $\hat H(1,\lambda,V,\theta)$ can be expressed explicitly
\begin{figure}
	\centering
	\includegraphics[width=0.48
	\textwidth]{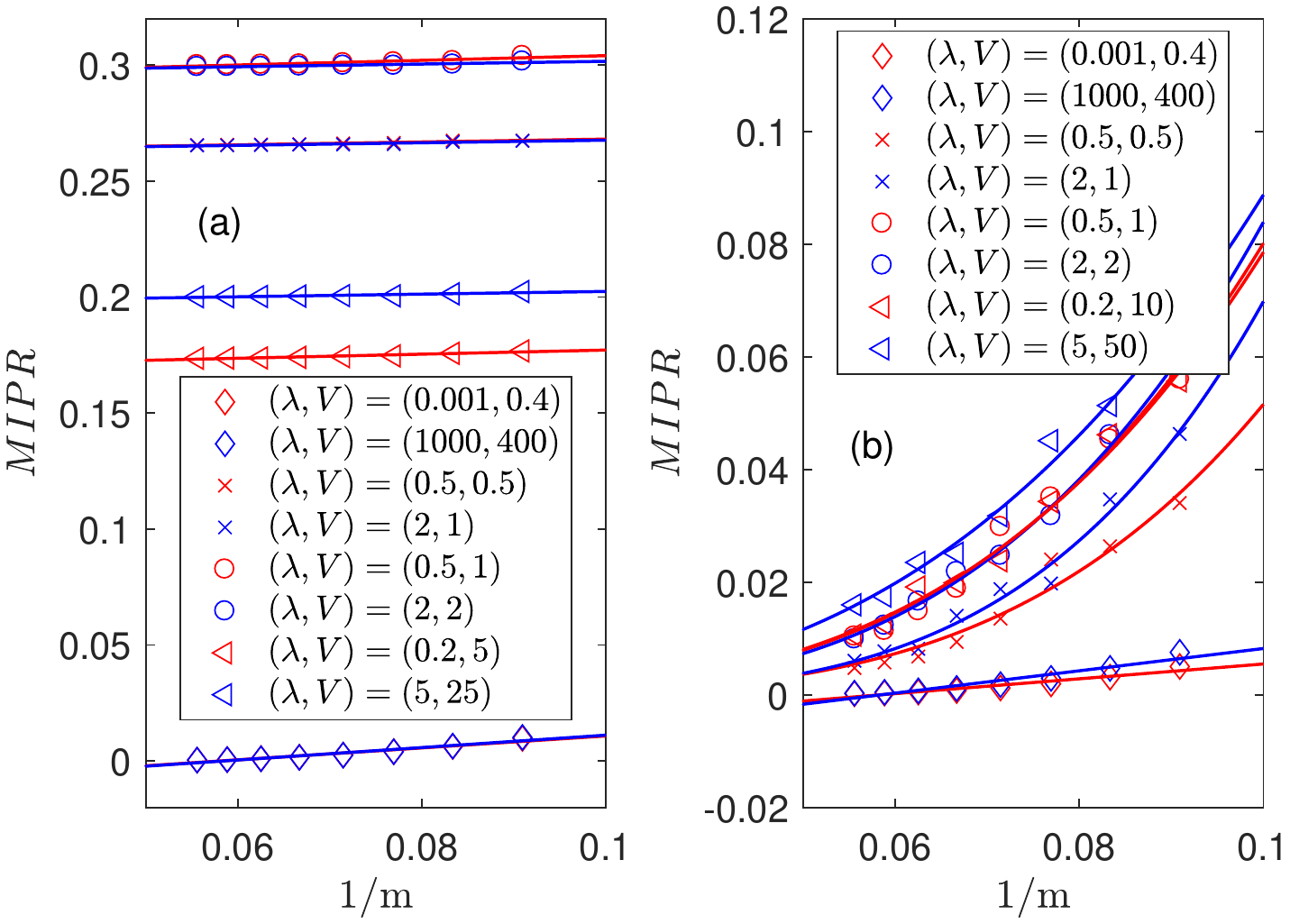}\\
	\caption{(Color online) (a) MIPR as a function of the inverse Fibonacci index $1/m$ for the various ($\lambda,V$) parameters in model I. (b) MIPR as a function of the inverse Fibonacci index $1/m$ for the various ($\lambda,V$) parameters in model II.}
	\label{fig2}
\end{figure}
\begin{equation}\label{difference1}
\begin{aligned}
&E a_n=(1+\lambda+V_{m1})b_{n-1}+(1-\lambda+V_{m2})b_{n},\\
&E b_n=(1-\lambda+V_{m3})a_{n}+(1+\lambda+V_{m4})a_{n+1},
\end{aligned}
\end{equation}
We exchange the positions of $1$ and $\lambda$, $\hat H(1,\lambda,V,\theta)$ become $\hat H(\lambda,1,V,\theta)$
\begin{equation}\label{difference2}
\begin{aligned}
&E a_n=(1+\lambda+V_{m1})b_{n-1}+(-1+\lambda+V_{m2})b_{n},\\
&E b_n=(-1+\lambda+V_{m3})a_{n}+(1+\lambda+V_{m4})a_{n+1},
\end{aligned}
\end{equation}
continue to transform Eq.~(\ref{difference2}), we can obtain
\begin{equation}\label{difference3}
\begin{aligned}
&E a_n=(1+\lambda+V_{m1})b_{n-1}+[1-\lambda+V_{m2}(+\pi)](-b_{n}),\\%
&E b_n=[1-\lambda+V_{m3}(+\pi)](-a_{n})+(1+\lambda+V_{m4})a_{n+1},
\end{aligned}
\end{equation}
Thus, $\hat H(\lambda,1,V,\theta)$ is equivalent to $\hat H(1,\lambda,V,\theta+\pi*\kappa)$, where $\kappa$ represents the parity of lattice dependence. Note the phase factor $\theta$ in $V_{n}=V \cos (2 \pi \alpha n+\theta)$ doesn't affect spectrum properties of the system, so we can ignore $\pi*\kappa$, and conclude $\hat H(\lambda,1,V,\theta)$ and $\hat H(1,\lambda,V,\theta)$ should have the same spectrum properties, i.e, localization properties. Consequently, Hamiltonian before exchange $\hat H(1,\lambda,V)$ has the same localization properties with Hamiltonian after exchange $\hat H(\lambda,1,V)$, i.e. $\frac{1}{\lambda}\hat H(1,\frac{1}{\lambda},\frac{V}{\lambda})$. Thus, the point $(\lambda,V)$ in the parameter space dually maps to the point $(\frac{1}{\lambda},\frac{V}{\lambda})$. The above steps can be summarized as follows
\begin{equation}\label{dual}
\hat H(1,\frac{1}{\lambda},\frac{V}{\lambda},\theta)\equiv\hat H(1,\lambda,V,\theta+\pi*\kappa)\leftrightarrow\hat H(1,\lambda,V,\theta).
\end{equation}

To support Eq.~(\ref{dual}), we implement numerical calculation of MIPR. The size of the system $L$ is chosen as the $m$th Fibonacci number $F_{m}$. The advantage of this arrangement is that the golden ratio can be approximately replaced by the ratio of two successive Fibonacci numbers, i.e.,
$\alpha=(\sqrt{5}-1)/2=\lim_{m\rightarrow\infty} F_{m-1}/F_{m}$. In Fig.~\ref{fig2} (a) we plot the trend of MIPR for dual points of model I as the system size increases. For $\{(\lambda,V)=(0.001,0.4),(1000,400)\}$ in extended phase, $\{(\lambda,V)=(0.5,0.5),(2,1)\}$ and $\{(\lambda,V)=(0.5,1),(2,2)\}$ in localized phase, MIPR of dual points conform to each other. However, for $\{(\lambda,V)=(0.2,5),(5,25)\}$ in localized phase, MIPR of dual points don't match up.

Here, we emphasize that the phase factor $\theta$ only does not affect the spectrum properties of the system, this does not mean that other physical quantities remain unchange. Regard to absolutely continuous spectrum (extended phase), the MIPR of $\hat H(1,\lambda,V,\theta)$ should tend to be consistent with the MIPR of $\hat H(1,\frac{1}{\lambda},\frac{V}{\lambda},\theta)$, due to both MIPR's being zero in the thermodynamic limit. However, Regard to point spectrum (localized phase), the MIPR of $\hat H(1,\lambda,V,\theta)$ should not necessarily tend to be consistent with the MIPR of $\hat H(1,\frac{1}{\lambda},\frac{V}{\lambda},\theta)$, due to both MIPR's just being nonzero in the thermodynamic limit, actually it just guarantees both MIPR's stay in the same order of magnitude.

In Fig.~\ref{fig2} (b), for $\{(\lambda,V)=(0.001,0.4),(1000,400)\}$ of model II in extended phase, the trend of MIPR also converges to zero. However, more interestingly, for the dual points in non-extended phase, the trend of MIPR in model II is not linearly extrapolated to a finite value like the localized phase in model I, distinctly, it obey the power-law curve $a*(\frac{1}{m})^b$. In the thermodynamic limit $m\rightarrow\infty$, the trend of MIPR approach zero, but the convergence rate is slower than that of the extended phase, this indicates that these dual points may belong to the critical phase.

From the point of view of dual-mapping, we can explicitly explain the relation of two red lines in Fig.~\ref{fig1} (c) and (d). The region $V\leq0.4(1-\lambda^2)$ is dually mapping to the region $V\leq0.4(\lambda-1/\lambda)$ for model I due to $(\lambda,V)\leftrightarrow(\frac{1}{\lambda},\frac{V}{\lambda})$. In the same way, the region $V\leq 1-\lambda$ is dually mapping to the region $V\leq \lambda-1$ for model II. Markedly, the rest regions for both model are the self-mapping region.

\section{quantum criticality}
\label{n3}
By comparing the value of MIPR of various V's for model I, it is very clearly that the self-mapping region of the model I is localized. However, The log trend of MIPR indicates the self-mapping region of the model II ($V> 1-\lambda$ and $V> \lambda-1$) is neither extended nor localized, but critical. However, it is difficult to study the localization nature of this model quantitatively. Here, we semi-analytically determines the emergence of quantum criticality in the self-mapping region of the model II.

In the first step, we prove that $V = 1$ is a quantum critical point when $\lambda = 0$. Thus, Eq.~(\ref{difference}) of model II can be expressed as
\begin{equation}\label{critical}
\begin{aligned}
E \psi_{n}=\left(1+V_{n-1}\right) \psi_{n-1}+\left(1+V_{n}\right) \psi_{n+1},
\end{aligned}
\end{equation}
where $V_{n}=V \cos (2 \pi \alpha n+\theta)$ denotes the off-diagonal quasiperiodic modulation.

We can study the localization properties of Eq.~(\ref{critical}) through the Lyapunov exponent (LE) \cite{Avila1}. To compute the LE $\gamma_0(E)$, we use Avila's global theory~\cite{Avila2} of quasiperiodic operators. Writing $T_n(\theta)$ the total transfer matrix of the model~(\ref{critical}),
the LE can be computed as~\cite{Avila1} $$\gamma_{\epsilon}(E)=\lim_{n\rightarrow \infty}\frac{1}{2 \pi n} \int_0^{2 \pi} \ln  \|T_n(\theta +i \epsilon)\| d\theta,$$ where $\|M\|$ represents the norm of the matrix $M$. The complexification of the phase ($\theta \rightarrow \theta+i \epsilon$) is crucial here, since our computation relies on Avila's global theory~\cite{Avila2}.
First note that the transfer matrix can be written as
\begin{equation*}
\begin{aligned}
T_{}(\theta)= \left(
\begin{array}{cc}
  \frac{E}{1+V_{n}} & -\frac{1+V_{n-1}}{1+V_{n}} \\
  1 & 0 \\
\end{array}
\right)=\frac{1}{1+V \cos (2 \pi \alpha n+\theta)}B_{}(\theta),\\
B_{}(\theta)= \left(
\begin{array}{cc}
  E & -1-V \cos [2 \pi \alpha (n-1)+\theta]\\
 1+V \cos (2 \pi \alpha n+\theta) & 0 \\
\end{array}
\right),
\end{aligned}
\end{equation*}
then LE can be expressed as
\begin{equation}
\begin{aligned}
\gamma_{\epsilon}(E)&=\lim_{n\rightarrow \infty}\frac{1}{n} \int \ln \|B_n(\theta +i \epsilon)\| d\theta\\
&+ \int \ln |\frac{1}{1+V \cos (2 \pi \alpha n+\theta+i \epsilon)}| d\theta\\
&=
\left\{
              \begin{array}{lr}
                 \gamma^{1}_{\epsilon}(E)-\ln\frac{1+\sqrt{1-V^2}}{2}~~~ if~~V<1 & \\
                 \gamma^{1}_{\epsilon}(E)-\ln|\frac{V}{2}|-2\pi\epsilon ~~~ if~~V\geq 1.
              \end{array}
    \right.
\end{aligned}
\end{equation}
where $\gamma^{1}_{\epsilon}(E)=\lim_{n\rightarrow \infty}\frac{1}{n} \int \ln  \|B_n(\theta +i \epsilon)\| d\theta$.
Let us then complexify the phase, and let $\epsilon$ goes to infinity, direct computation of $B(\theta+i\epsilon)$ yields
\begin{equation}
B(\theta+i\epsilon)=\frac{e^{2\pi\epsilon}e^{-i2\pi(\theta-\frac{1}{2}\alpha)}}{2}\left(
\begin{array}{cc}
 0 &   -Ve^{i \pi \alpha} \\
Ve^{-i \pi \alpha} & 0 \\
\end{array}
\right) + o(1).
\notag
\end{equation} Thus we have
$\gamma^{1}_{\epsilon}(E)=2\pi\epsilon+\log|\frac{V}{2}| +o(1).$
By the extended A. Avila's global theory to jacobi matrix~\cite{Jitomirskaya1,Jitomirskaya2},
$\gamma^{1}_{\epsilon}(E)=2\pi\epsilon+\log|\frac{V}{2}|.$
Consequently, we obtain
\begin{equation}
\gamma_{\epsilon}(E)=
\left\{
              \begin{array}{lr}
                \log| \frac{V}{1+\sqrt{1-V^2}}|+2\pi\epsilon~~~ if~~V<1 & \\
                 0 ~~~~~~~~~~~~~~~~~~~~~~~~~~~~ if~~V\geq 1.
              \end{array}
\right.
\end{equation}
Thus, if the energy $E$ lies in the spectrum, we have $\gamma_{0}(E)= \ln| \frac{V}{1+\sqrt{1-V^2}}|$ or $0$.
If $V < 1$, we have $\ln| \frac{V}{1+\sqrt{1-V^2}}|<0$, in this place Avila dubbed it as subcrtical, and he has prove there is only absolutely continuous spectrum~\cite{Avila3,Avila4}.
If $V \geq 1$, $\frac{1}{1+V \cos (2 \pi \alpha n+\theta)}$ always has singularity, by the paper~\cite{Jitomirskaya2}, we know there is There is no absolutely continuous spectrum. At the same time, through the paper~\cite{Han}, we can know that there is no point spectrum. Therefore in this case, we only singular continuous spectrum. Consequently, we analytically demonstrate the emergence of quantum criticality in the self-mapping region ($\lambda=0$ and $V \geq 1$).

In the second step, we prove that $\lambda = 1$ is the degeneracy point of energy level when $V = 0$. Thus, Eq.~(\ref{difference}) of the Hamiltonian reduces to the SSH model
\begin{equation}\label{ssh}
\begin{aligned}
&E a_n=(1+\lambda)b_{n-1}+(1-\lambda)b_{n},\\
&E b_n=(1-\lambda)a_{n}+(1+\lambda)a_{n+1},
\end{aligned}
\end{equation}
When $\lambda = 1$, Eq.~(\ref{ssh}) become
\begin{equation}\label{ssh1}
\begin{aligned}
&E a_n=2 b_{n-1},\\
&E b_n=2 a_{n+1},
\end{aligned}
\end{equation}
Thus, we obtain a degenerate energy spectrum $E=2$. Here, $\lambda = 1$ is not conventional quantum critical point, this is no topological phase transition. However, the degeneracy of energy levels are commonly accompanied by the emergence of quantum criticality.

Reference~\cite{Kinross} studies the phase diagram of the transverse field Ising model under both the transverse magnetic field and the finite temperature and experimentally confirms the quantum critical behavior. In that model, quantum fluctuations controlled by the nonthermal parameter $g$ lead to a phase transition at a critical value $g_c$, the so called quantum critical point, already at zero temperature. And there exists a thermal phase transition at a critical temperature $T_c$. The interplay of quantum fluctuations $g_c$ and thermal fluctuations $T_c$ opens up a progressively broader, $V$-shaped quantum critical region extending much above the zero temperature.
Analogically, under the critical dimerization strength $\lambda_c=1$ and the critical quasiperiodic modulation strength $V_c=1$ in our model, a similar shaped quantum critical region can emerge, thus we explain the emergence of quantum criticality of the model II qualitatively. 

\begin{figure}[t]
    \centering
    \includegraphics[width=0.48\textwidth]{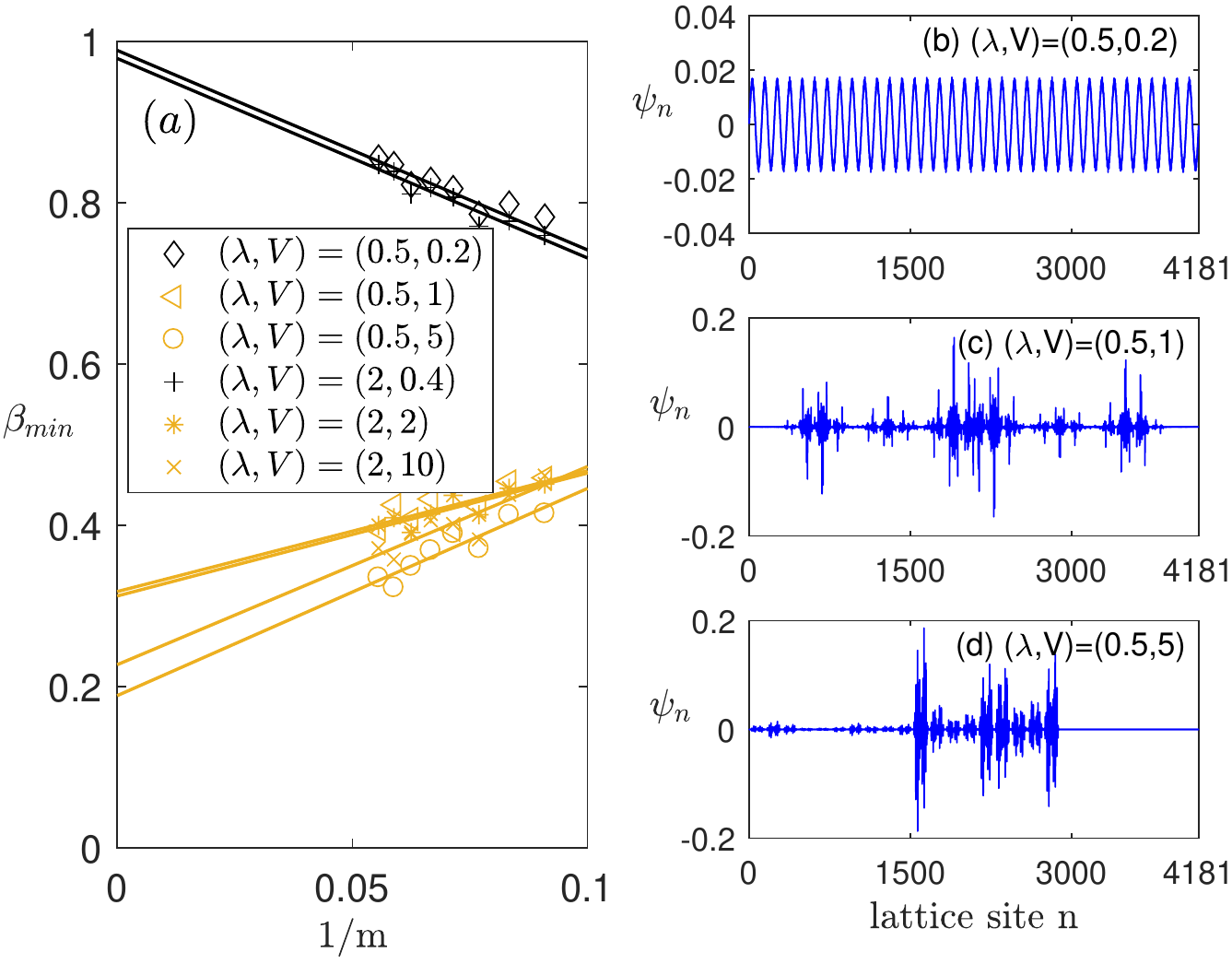}
    \caption{(a) $\beta_{min}$ as a function of the inverse Fibonacci index $1/m$ for the various ($\lambda,V$) parameters in model II. The brown markers correspond to the critical phase, and the black markers correspond to the extended phase. (b) the representative wave function in $(\lambda,V)=(0.5,2)$ is extended. (c) the representative wave function in $(\lambda,V)=(0.5,1)$ is critical. (d) the representative wave function in $(\lambda,V)=(0.5,5)$ is critical. The system size is set to be $L=F_{m} = 4181$.}
     \label{fig3}
\end{figure}

To characterize the wave functions and validate our conjecture of critical states, we use the multifractal theory, which has been widely applied to standard quasiperiodic models, such as the Aubry-Andr\'e model~\cite{fractal}.
For each wave function $\psi_n^j=(a_n^j,b_n^j)^T$, a scaling exponent $\beta_{n}^j$ can be extracted from the $n$th
on-site probability $P_{n}^j = \vert\psi_n^j \vert^2 \sim (1/F_{m})^{\beta_{n}^j}$.
According to the multifractal theorem, when the wave functions are extended, the maximum of $P_{n}^j$ scales as $\max(P_{n}^j)
\sim (1/F_{m})^1$, i.e., $\beta_{min}^j=\min(\beta_{n}^j)=1$. On the other hand, when the wave functions are localized, $P_{n}^j$ peaks at very
few sites and nearly zero at the other sites, yielding $\max(P_{n}^j) \sim (1/F_{m})^0$ and $\beta_{min}^j=\min(\beta_{n}^j)=0$. As for the critical
wave functions, the corresponding $\beta_{min}^j$ is located within the interval $\left(0,~1\right)$, and can be used to discriminate extended and critical states.
It should be noted, however, that $\beta_{min}^j$ for the critical
Aubry-Andr\'{e} model fluctuates sharply with system size $L$.
In order to reduce finite-size effects, we thus use the average of $\beta_{min}^j$, i.e. $\beta_{min}=\frac{1}{2L}\sum^{2L}_{j=1}\beta_{min}^j$ over wave functions of the system so as to distinguish the critical state in practical numerical calculations.

Figure \ref{fig3}(a) plots $\beta_{min}$ as a function of the inverse Fibonacci index $1/m$. It clearly shows that $\beta_{min}$ is between $0$ and $1$ in the large $L$ limit for the self-mapping region $\{(\lambda,V)=(0.5,1),(2,2),(0.5,5),(2,10)\}$, hence suggesting that wave functions in the self-mapping region are critical.
Conversely, for $\{(\lambda,V)=(0.5,0.2),(2,0.4)\}$, $\beta_{min}$ asymptotically tends to 1 in the thermodynamic limit, indicating that the corresponding wave functions are extended. Figure~\ref{fig3} [panels~(b), (c), and (d)] plots the representative wave functions of the extended and self-mapping region, respectively. An inspection of the figure clearly indicates that the wave function is extended for $\{(\lambda,V)=(0.5,0.2)\}$ [panels~(b)].
In contrast, the wave functions of $\{(\lambda,V)=(0.5,1),(0.5,5)\}$ in the self-mapping region [panels~(c) and (d)] are neither localized nor extended over the whole space. Instead, they display clear self-similarities, which is the characteristic of critical states.

\section{Conclusion}
\label{n4}
In summary, We study two off-diagonal quasiperiodic models and unveil a general dual-mapping relation
in parameter space of the dimerization strength $\lambda$ and the quasiperiodic modulation strength $V$.
Moreover, we demonstrated semi-analytically and numerically that a specific off-diagonal quasiperiodic model can host a broad quantum critical region.
This is different from the customary diagonal quasiperiodic models, in which quantum criticality can only emerge at the quantum phase transition point.
Our discovery will enrich the abundant localization phenomena in the quasiperiodic systems.

\begin{acknowledgments}
Tong Liu thank Pei Wang for fruitful discussions. Tong Liu acknowledges Natural Science Foundation of Jiangsu
Province (Grant No. BK20200737) and NUPTSF (Grant No. NY220090 and No. NY220208). X. Xia is supported by Nankai Zhide Foundation.
\end{acknowledgments}

\end{document}